# Exact combinatorial approach to finite coagulating systems through recursive equations

Michał Łepek*, Paweł Kukliński, Agata Fronczak, Piotr Fronczak

Warsaw University of Technology, The Faculty of Physics, Koszykowa 75, 00-662, Warsaw, Poland

*lepek@if.pw.edu.pl

**Abstract**

The work outlines an exact combinatorial approach to finite coagulating systems. In the classic approach one uses the mean-field Smoluchowski coagulation. However, the assumptions of the mean-field theory are often unmet in real systems which causes that the accuracy of the solution is limited. In our approach, cluster sizes and time are discrete, and the binary aggregation alone governs the time evolution of the systems. By considering the growth histories of all possible clusters, the exact expression is derived for the probability of a coagulating system with an arbitrary kernel being found in a given cluster configuration when monodisperse initial conditions are applied. Then, the average number of such clusters and that average's standard deviation can be calculated. In this work, recursive equations for all possible growth histories of clusters are introduced. The correctness of our expressions is proved based on the comparison with numerical results obtained for systems with the constant, multiplicative and additive kernels, the two latter for the first time using this framework. In addition, the results obtained are compared with the results arising from the solutions to the mean-field Smoluchowski equation which are outperformed by our theoretical predictions.

## I. Introduction

Coagulation processes (also called as aggregation, coalescence, etc.) are very common in nature. They are not only of great interest in pure sciences, inter alia, physics [1–3], chemistry [4–6], biology [7] or mathematics [8–10] but also underlie everyday–life phenomena which can be: blood coagulation, milk curdling, cloud forming. Coagulation is the basis for wide variety of technological applications which are, among others, food processing, road monitoring, clinical diagnosis or water treatment [11].

The simplest coagulation process can be regarded as evolution of a closed system of clusters that merge irreversibly in effect of binary collisions (coagulation acts), according to the scheme:

$$(i) + (j) \xrightarrow{K(i,j)} (i + j) \tag{1}$$

where $(i)$ stands for a cluster of mass $i$ and $K(i,j)$ is the rate of the process which is called the coagulation kernel. The number of clusters decreases in time and eventually all of the clusters join into one single cluster.

The most–widely known approach to modeling coagulation process is based on the Smoluchowski coagulation equation [12] which requires several assumptions. These are: infinite size of the system studied and continuous (instead of discreet) cluster sizes. These conditions are clearly unmet in small systems. They also cause problems in large systems, for long times of system evolution, when the number of clusters becomes smaller and smaller. , Another weakness of the Smoluchowski equation appears when it is applied to the so–called gelling kernels. It does not model the process in well, for instance, in the multiplicative kernel it fails to conserve the mass after a finite time of the gelation point. In addition, the solutions arising from the Smoluchowski equation are stochastically incomplete and describe only the average behavior of clusters, not providing any information on deviations from it. On the other hand, a good point of the approach based on this equation is that explicit analytical solutions are known for particular kernels, e.g. constant $(K(i,j) = const)$, multiplicative $(K(i,j) = ij)$ and additive $(K(i,j) = i + j)$, and there is some literature on the existence and uniqueness of solutions to some general classes of kernels [13-19].

Given the problems with the Smoluchowski coagulation equation, a new stochastic approach to finite coagulating systems has been proposed in opposition to deterministic mean–field and infinite–volume Smoluchowski approach [20–24]. Other important contributions were given by Lushnikov, including more effective theoretical analysis of the coagulating systems with constant and multiplicative kernels [25–27].

In this paper, we develop the approach proposed previously in [28] where the assumptions are that time is discrete and one coagulation act occurs in each time step. Successive steps define then the space of available states and, by studying possible growth histories of clusters using combinatorial expressions, the probability distribution over the state space is determined. This approach has been proved to be effective for the system with constant kernel and monodisperse initial conditions. What we improve in this work, is introducing more general method for calculating the number of possible histories of a cluster and proving its effectiveness for constant, additive and multiplicative kernels.

The paper is organized as follows. In Section II basics of the combinatorial approach are briefly presented. Section III provides detailed description of our method for calculating the number of possible internal states of a cluster for different kernels. Section IV covers probability distribution and cluster statistics. The results of theoretical predictions in comparison to the numerical simulations are provided in Section V. In Section VI concluding remarks and possible extensions to this work are stated.

## II. Combinatorial approach

The combinatorial approach used here  was first introduced in [28]. When investigating the aggregating system, it is vital to start with some basic observations. Assuming discrete time and monodisperse initial conditions (all of the clusters are monomers, i.e. of size of 1 unit) and if a single coagulation act occurs in one time step, then the total number of clusters, $k$, at time $t$ is

$$k = N - t, \tag{2}$$

where $N$ is the total number of monomeric units in the system. $N$ is equivalent to the initial number of clusters as the number of monomeric units in the system do not change during the evolution of the system (preservation of mass). The state of the system at time $t$ is described by

$$\Omega(t) = \{n_1, n_2, \dots, n_g, \dots, n_N\}, \tag{3}$$

where $n_g \geq 0$ stands for the number of clusters of mass $g$, therefore $g$ is the number of monomeric units included in the cluster and $n_1$ corresponds to monomers, $n_2$ – dimers, $n_3$ – trimers and so on. It is clear that during the coagulation process the sequence $\{n_g\}$ is not arbitrary but satisfies:

$$\sum_{g=1}^{N} n_g = k \qquad \text{and} \qquad \sum_{g=1}^{N} g\, n_g = N, \tag{4}$$

which corresponds to the preservation of the number of monomeric units in the system. The total number of states of the coagulation process, depending on time $t$, can be described using the Stirling number of the second kind which gives the number of ways to partition a set of $N$ objects into $k$ subsets:

$$\overline{\Omega}(t) = S\big(N, k(t)\big) \tag{5}$$

although this information is not sufficient to find probability distribution over the state space $P(\Omega)$, because these states are not equiprobable.As further shown in the paper, an useful tool for analyzing aggregation phenomena are partial (also called as incomplete) exponential Bell polynomials (Bell polynomials for short). Bell polynomial are defined as:

$$B_{N,k}(x_1, x_2, \ldots, x_{N-k+1}) = B_{N,k}(\{x_g\}) = N! \sum_{\{n_g\}} \prod_{g=1}^{N-k+1} \frac{1}{n_g!} \left(\frac{x_g}{g!}\right)^{n_g} \tag{6}$$

where the summation is taken over all non-negative integers $\{n_g\}$ that satisfy Eq. (4). Using Bell polynomials one can obtain a detailed information about the partition of an arbitrary set. For example, considering a set of $N = 6$ monomeric units and $k = 3$ clusters we instantaneously have: $B_{6,3}(x_1, x_2, x_3, x_4) = 15x_1^2 x_4 + 60x_1 x_2 x_3 + 15x_2^3$, which encodes the information that there are 15 ways to partition a set of 6 as $1 + 1 + 1$, 60 ways to partition it as $1 + 2 + 3$, and 15 ways to partition as $2 + 2 + 2$, therefore the total number of possible partitions is consistent with the result given by the Stirling number: $S(6,3) = 90$.

There are two sources of combinatorial equations for coagulating systems. The first one is related to partitioning the system into subsets. Let $N$ distinguishable objects be into $k$ non-empty and disjoint subsets of $c_i > 0$ elements each, where $\sum_{i=1}^{k} c_i = N$. As we can choose $c_1$ in $\binom{N}{c_1}$ ways, $c_2$ in $\binom{N - c_1}{c_2}$ ways, there are

$$\begin{aligned} \binom{N}{c_1, c_2, \ldots, c_k} &= \binom{N}{c_1}\binom{N - c_1}{c_2} \ldots \binom{N - (c_1 + c_2 + \cdots + c_{k-1})}{c_k} = N! \prod_{i=1}^{k} \frac{1}{c_i!} \\ &= N! \prod_{g=1}^{N-k+1} \left(\frac{1}{g!}\right)^{n_g} \end{aligned} \tag{7}$$

of such partitions; $n_g \geq 0$ stands for the number of subsets of size $g$, with the largest subset size equal to $N - k + 1$. The change of the multiplication index $i$ to $g$ means that we no longer multiply over the subsets (clusters) but over the sets of subsets (clusters) of a given size determined by $g$. If in such a composition each of $n_g$ subsets (clusters) of size $g$ can be in any of $x_g \geq 0$ internal states and the order of clusters does not matter (division by $n_g!$) then the number of partitions becomes

$$N! \prod_{g=1}^{N-k+1} \frac{1}{n_g!} \left(\frac{x_g}{g!}\right)^{n_g}. \tag{8}$$

Summing (8) over all integers $\{n_g\}$ satisfying the constraints specified by Eq. (4) one gets the partial Bell polynomial $B_{N,k}(\{x_g\})$. Internal state $x_g$ of the cluster corresponds to the number of ways that the cluster could be created (other words, the number of possible histories of the cluster). In general, this variable varies for different kernels and the methodology for obtaining it is presented in Sec. III. Eq. (8) gives the number of ways of partition of $N$ monomers into $k$ clusters for a given state $\Omega$ of the system. However, Eq. (8) also entails implicit assumption that all the clusters in the system arise at once, simultaneously, in the same time step. This is an obvious untruth as only one coagulation act is allowed to occur per one time step. Additionally, coagulation acts corresponding to different clusters may alternate with each other increasing the number of ways a given microstate can be created. Although the individual evolution of each cluster is covered by the sequence $\{x_g\}$ as far as here we did not take into account that coagulation acts leading to create the cluster can occur in different steps in time. In other words, a given state can arise as a result of different sequences of intermixed coagulation acts corresponding to different clusters. Therefore, the distribution of coagulation acts in time is the second source of combinatorial equations.

As mentioned before, each cluster of size $g$ requires $g-1$ acts of coagulation, the process starts from the monodisperse initial state and the total number of coagulation acts is equal to $t$. Thus, as each dimer requires one step in time to be created we can choose the timestep for creating the first dimer in $\binom{t}{1}$ ways because we use one of $t$ available timesteps. Then we can choose the timestep for the second dimer in $\binom{t-1}{1}$ ways, for the third dimer in $\binom{t-2}{1}$ ways and so on until for the last dimer we have $\binom{t-n_2+1}{1}$ ways as $n_2$ is the number of dimers. We now can perform the reasoning again for trimers keeping in mind that trimers require two timesteps to be created and that available space of timesteps is now decreased. The overall number of sequences in time corresponding to each of possible microscopic realizations (8) of the system is then given by:

$$\left[\binom{t}{1}\binom{t-1}{1}\dots\binom{t-n_2+1}{1}\right]\left[\binom{t-n_2}{2}\binom{t-n_2-2}{2}\dots\binom{t-n_2-2(n_3-1)}{2}\right]\dots \\ = \frac{t!}{(1!)^{n_2}(2!)^{n_3}\dots\left((g-1)!\right)^{n_g}\dots} = t!\prod_{g=1}^{N-k+1}\frac{1}{\left((g-1)!\right)^{n_g}} \tag{9}$$

By multiplying (8) and (9) one gets the number of ways that the system state can be created. Assuming that these events are just as likely, $W(\Omega)$ also defines the thermodynamic probability that the system will be found in $\Omega$.

## III. Calculating the number of all possible internal states of a cluster, $x_g$, through recursive equations

### A. Constant kernel

As shown in [28], in the case of constant kernel the number of all possible internal states of a cluster of size $g$ (all possible histories of its growth) can be calculated in a very simple way. As the reaction rate is constant and the probability of coalescing two smaller clusters into bigger one is the same during the evolution, in the first time step two monomers can be chosen out of $g$ available monomers and merged. In the second time step two clusters out of $g-1$ available clusters are chosen and merged. In the third time step the next two clusters available out of $g-2$ are merged, and so on. The above can be written as

$$x_g = \binom{g}{2}\binom{g-1}{2}\binom{g-2}{2}\dots\binom{2}{2} = \frac{g!\,(g-1)!}{2^{g-1}}. \tag{10}$$

However, this simple reasoning cannot be extended to cover other kernels, e.g. multiplicative and additive kernels. To solve this problem we propose more general methodology of recursive equations for $x_g$ which can be used, as we show in the later part of this work, to derive $x_g$ for multiplicative and additive kernels. Bearing in mind that a cluster of size $g$ is created by merging clusters of sizes $k$ and $(g-k)$, we put

$$x_g = \frac{1}{2}\sum_{k=1}^{g-1}\binom{g}{k}\binom{g-2}{k-1}x_k x_{g-k} \tag{11}$$

where $x_k$ is the number of possible internal states (the number of ways of creation) of cluster of size $k$ and $x_{g-k}$ is the number of possible internal states of cluster of size $(g-k)$. The first Newton symbol $\binom{g}{k}$ calculates ways of choosing clusters of size $k$ out of $g$ monomers as we can divide the cluster of size $g$ into subclusters of size $k$ and size $(g-k)$ in exactly $\binom{g}{k}$ ways. The second Newton symbol $\binom{g-2}{k-1}$ stands for the fact that the coagulations acts of clusters of sizes $(g-k)$ and $k$ could appear in different steps in time. The sense of this symbol is analogous as in Eq. (10). Coagulations acts related to creation of cluster of size $k$ could occur in $k-1$ time steps out of the total number of $g-2$ time steps needed to create clusters of sizes $k$ and $(g-k)$. The sum is calculated over the allowed size of the merged cluster and the factor of $\frac{1}{2}$ is used to prevent double counting of coagulations acts that could result in creating cluster of size $g$.

We will now prove that one can derive (10) from (11). After expanding Newton symbols we have:

$$x_g = \frac{1}{2}\sum_{k=1}^{g-1}\frac{g!}{k!\,(g-k)!}\frac{(g-2)!}{(k-1)!\,(g-k-1)!}x_k x_{g-k}\,. \tag{12}$$

Now, we introduce a new parameter

$$y_g = \frac{x_g}{g!\,(g-1)!} \tag{13}$$

After elementary transformations Eq. (12) can be written as follows:

$$(g-1)y_g = \frac{1}{2}\sum_{k=1}^{g-1}y_k y_{g-k}\,. \tag{14}$$

We now multiply both sides of (14) by $\sum_{g=1}^{\infty} z^g$ and have

$$\sum_{g=1}^{\infty}(g-1)y_g z^g = \frac{1}{2}\sum_{g=1}^{\infty}\sum_{k=1}^{g-1}(y_k z^k)\left(y_{g-k}z^{g-k}\right). \tag{15}$$

Transforming left–hand side gives:

$$\begin{aligned}\sum_{g=1}^{\infty}(g-1)y_g z^g &= \sum_{g=1}^{\infty} g y_g z^g - \sum_{g=1}^{\infty} y_g z^g = z\sum_{g=1}^{\infty} g y_g z^{g-1} - G(z)\\ &= z\frac{\partial}{\partial z}\sum_{g=1}^{\infty} y_g z^g - G(z) = z\frac{\partial}{\partial z}G(z) - G(z)\end{aligned} \tag{16}$$

where

$$G(z) \equiv \sum_{g=1}^{\infty} y_g z^g \tag{17}$$

is the generating function for $y_g$. Similarly, transforming right–hand side of Eq. (15) we have

$$\begin{aligned}\frac{1}{2}\sum_{g=1}^{\infty}\sum_{k=1}^{g-1}(y_k z^k)\left(y_{g-k}z^{g-k}\right) &= \frac{1}{2}\left(\sum_{k=1}^{\infty} y_k z^k\right)\left(\sum_{g-k=1}^{\infty} y_{g-k}z^{g-k}\right)\\ &= \frac{1}{2}G(z)G(z)\end{aligned} \tag{18}$$

The change in summation boundaries is due to the observation that in case of $y_k z^k$ the sum takes all expressions from 1 to infinity and in case of $y_{g-k}z^{g-k}$ the sum takes all expressions from $g-k=1$ to infinity.

The above transformations lead us to the differential equation for $G(z)$:

$$z\frac{\partial}{\partial z}G(z) - G(z) = \frac{1}{2}\left(G(z)\right)^2, \tag{19}$$

which is an ordinary differential equation with separated variables and its solution has the form:

$$G(z) = \frac{2Az}{1-Az} = 2Az\left(\frac{1}{1-Az}\right) = 2Az\sum_{g=0}^{\infty}(Az)^g = 2\sum_{g=1}^{\infty}(Az)^g \tag{20}$$

As $G(z) = \sum_{g=1}^{\infty} y_g z^g$ , see Eqs. (13) and (17), for the first elements of series we have,

$$2A^g = \frac{x_g}{g!\,(g-1)!} \tag{21}$$

As we know that $x_1 = 1$, we can calculate the constant, $A = \frac{1}{2}$, and obtain

$$x_g = \frac{g!\,(g-1)!}{2^{g-1}} \tag{22}$$

which remains in compliance with (10).

**B. Multiplicative kernel**

For the multiplicative kernel the probability of a coagulation act is proportional to the product of masses of coagulating clusters. To take into account this feature we must modify Eq. (11) to the form of:

$$x_g = \frac{1}{2}\sum_{k=1}^{g-1}\binom{g}{k}\binom{g-2}{k-1}x_k x_{g-k}\cdot k(g-k) \tag{23}$$

where the latter coefficient $k(g-k)$ modifies $x_g$ in accordance to the masses of two contributing sub-clusters. For example, if one of sub-clusters is a monomer with $k=1$ then bigger sub-cluster of mass $(g-k)$ only alternates $x_g$.

After expanding Newton symbols and this time substituting $y_g = \frac{x_g}{(g-1)!(g-1)!}$, we have:

$$(g-1)y_g = \frac{1}{2}g\sum_{k=1}^{g-1}y_k y_{g-k} \tag{24}$$

Applying analogous steps and transformations as for the constant kernel, one obtains the equation for the generating function $G(z)$ (see Eq. (17)) for the multiplicative kernel:

$$G(z)\cdot e^{-G(z)} = Az \tag{25}$$

The Eq. (25) has a form of $f(G)=F$ where $f(G)$ stands for the left–hand side of Eq. (25) and $F$ stands for the right–hand side of Eq. (25). Having the equation of that form and applying the Lagrange inversion (see p. 148 in [1]), one can obtain the series representation of the inverse function $G(F)$:

$$G(F) = \sum_{n\geq 1}\frac{n^{n-1}}{n!}F^n \tag{26}$$

which is in our case:

$$G(z) = \sum_{g=1}^{\infty}y_g z^g = \sum_{g=1}^{\infty}\frac{g^{g-1}}{g!}(Az)^g. \tag{27}$$

From the initial condition, $y_1 = 1$, we can have the constant $A=1$. Considering the expressions from under the sums and going back from the substitution $y_g$ we have:

$$y_g = \frac{g^{g-1}}{g!} = \frac{x_g}{(g-1)!\,(g-1)!}. \tag{28}$$

Finally, the number of ways of creating the cluster of size $g$ for the multiplicative kernel is

$$x_g = (g-1)! \cdot g^{g-2}\,. \tag{29}$$

**C. Additive kernel**

In case of the additive kernel the probability of the coagulation act is proportional to the sum of masses of coagulating clusters. Modifying Eq. (11) to meet this criterion, we have:

$$\begin{aligned} x_g &= \frac{1}{2}\sum_{k=1}^{g-1}\binom{g}{k}\binom{g-2}{k-1}x_k x_{g-k}\cdot\big(k+(g-k)\big) \\ &= \frac{1}{2}\sum_{k=1}^{g-1}\binom{g}{k}\binom{g-2}{k-1}x_k x_{g-k}\cdot g \end{aligned} \tag{30}$$

If we expand Newton symbols and substitute $y_g = \frac{x_g}{g!(g-1)!}$, we obtain:

$$(g-1)y_g = \frac{1}{2}g\sum_{k=1}^{g-1} y_k y_{g-k} \tag{31}$$

which is, fortunately, exactly the same equation as Eq. (24) with the only difference in defining $y_g$. Therefore, the solution is identical and in this case we obtain:

$$y_g = \frac{g^{g-1}}{g!} = \frac{x_g}{g!\,(g-1)!} \tag{32}$$

Finally, for the multiplicative kernel the number of ways of creating the cluster of size $g$ is

$$x_g = (g-1)! \cdot g^{g-1}\,. \tag{33}$$

**IV. Probability distribution over the state space and cluster statistics**

The main aim of the analysis is to derive the probability distribution over the state space of the system, i.e. the probability, $P(\Omega)$, of a coagulating system being found in a given state $\Omega$. Non-equilibrium character of the process results that the allowed states of the system are not equiprobable and $P(\Omega) \neq \overline{\Omega}(\mathrm{t})^{-1}$. Having found the thermodynamic probabilities, $W(\Omega)$, one can easily have [28]:

$$P(\Omega) = \frac{W(\Omega)}{Z} \tag{34}$$

where $Z = \sum_{\Omega} W(\Omega)$ is the normalizing factor. Due to the results of previous sections, we can write down the thermodynamic probabilities in the following way:

$$W(\Omega) = \left[ t! \prod_{g=1}^{N-k+1} \frac{1}{\left((g-1)!\right)^{n_g}} \right] \left[ N! \prod_{g=1}^{N-k+1} \frac{1}{n_g!} \left( \frac{x_g}{g!} \right)^{n_g} \right]$$
$$= t!\, N! \prod_{g=1}^{N-k+1} \frac{1}{n_g!} \left( \frac{x_g}{(g-1)!\, g!} \right)^{n_g} \tag{35}$$

and calculate the parameter $Z$ over all states of the system using Bell polynomials to simplify the expression:

$$Z = \sum_{\Omega} W(\Omega) = \ t! \left[ N! \prod_{g=1}^{N-k+1} \frac{1}{n_g!} \left( \frac{x_g}{(g-1)!\, g!} \right)^{n_g} \right] = t!\, B_{N,k}\left( \left\{ \frac{x_g}{(g-1)!} \right\} \right). \tag{36}$$

For further simplification let us now have $\omega_g = \frac{x_g}{(g-1)!}$ and thus

$$Z = t!\, B_{N,k}\left( \{ \omega_g \} \right). \tag{37}$$

The probability distribution is then specified as

$$P(\Omega) = \frac{W(\Omega)}{Z} = \frac{N!}{B_{N,k}\left( \{ \omega_g \} \right)} \prod_{g=1}^{N-k+1} \frac{1}{n_g!} \left( \frac{\omega_g}{g!} \right)^{n_g} \tag{38}$$

and provides the most detailed information about the finite–size coalescing system.

Having the probability distribution of clusters calculated, one can straightforwardly proceed to determine the average number of clusters of a given size and the standard deviation of that average. It has been shown in [28] that the average number of clusters of a given size, $n_s$, can be find as

$$\langle n_s \rangle = \sum_{\Omega} n_s(\Omega) P(\Omega) = \binom{N}{s} \omega_s \frac{B_{N-s,k-1}\left( \{ \omega_g \} \right)}{B_{N,k}\left( \{ \omega_g \} \right)} \tag{39}$$

and the corresponding standard deviation of this average as

$$\sigma_s = \sqrt{\langle n_s (n_s - 1) \rangle + \langle n_s \rangle - \langle n_s \rangle^2} \tag{40}$$

where

$$\langle n_s (n_s - 1) \rangle = \binom{N}{s,\, s} {\omega_s}^2 \frac{B_{N-2s,k-2}\left( \{ \omega_g \} \right)}{B_{N,k}\left( \{ \omega_g \} \right)}. \tag{41}$$

The average number of clusters of a given size for the constant kernel, as known before (see Eq. (38) in [28]), is given by

$$\langle n_s \rangle_c = k \frac{\binom{N-1-s}{k-2}}{\binom{N-1}{k-1}}. \quad (42)$$

As newly found, the average number of clusters of a given size for the multiplicative and additive kernels are respectively given by

$$\langle n_s \rangle_m = \binom{N}{s} s^{s-2} \frac{B_{N-s,k-1}(\{g^{g-2}\})}{B_{N,k}(\{g^{g-2}\})} \quad (43)$$

$$\langle n_s \rangle_a = \binom{N}{s} s^{s-1} \frac{B_{N-s,k-1}(\{g^{g-1}\})}{B_{N,k}(\{g^{g-1}\})}. \quad (44)$$

The latter one can be simplified, due to the formula $B_{N,k}(\{g^{g-1}\}) = \binom{N-1}{k-1} N^{N-k}$ , to the form of

$$\langle n_s \rangle_a = \binom{N}{s} s^{s-1} \frac{\binom{N-1-s}{k-2}(N-s)^{N-s-k+1}}{\binom{N-1}{k-1} N^{N-k}} \quad (45)$$

which does not consist of Bell polynomials.

## V. Results

In this chapter, we present the results of our theoretical predictions compared to the results of numerical simulations and the solutions that arise from the Smoluchowski equations. The discrete–time forms of the latter ones are:

$$c_k(t) = \frac{t^{k-1}}{(1+t)^{k+1}} \quad (46)$$

$$c_k(t) = \frac{k^{k-2}}{k!} t^{k-1} e^{-kt} \quad (47)$$

$$c_k(t) = \frac{e^{-t}}{k!} e^{(1-e^{-t})k} \big((1-e^{-t})k\big)^{k-1} \quad (48)$$

for the constant, multiplicative and additive [1, 10] kernels, respectively. The solutions from the Smoluchowski equations are concentrations of clusters of size $k$, and need to be multiplied by the total number of monomers in the system, $N$, to obtain average number of clusters as $c_k(t) = \lim_{N\to\infty} \frac{\langle n_k(t) \rangle}{N}$. In addition, the time in Eqs. (46–48) is defined in the other way than in the exact approach and $t_s = t/(2N)$, where $t_s$ – time in Eqs. (46–48), $t$ – time in our approach.

The comparison of our theoretical predictions and simulation results are presented in Fig. 1 for the constant, multiplicative and additive kernels. In numerical simulations different kernels were represented as different schemes of choosing clusters to be merged. In case of the constant kernel, we randomly choose two clusters at each time step. In case of the multiplicative kernel, the probability of choosing a specific cluster is proportional to its size. In case of the additive kernel, one cluster is chosen randomly from the set of clusters while the second one is chosen with probability which is proportional to its size. The additive kernel can be simulated in such a way as can be seen in [29].

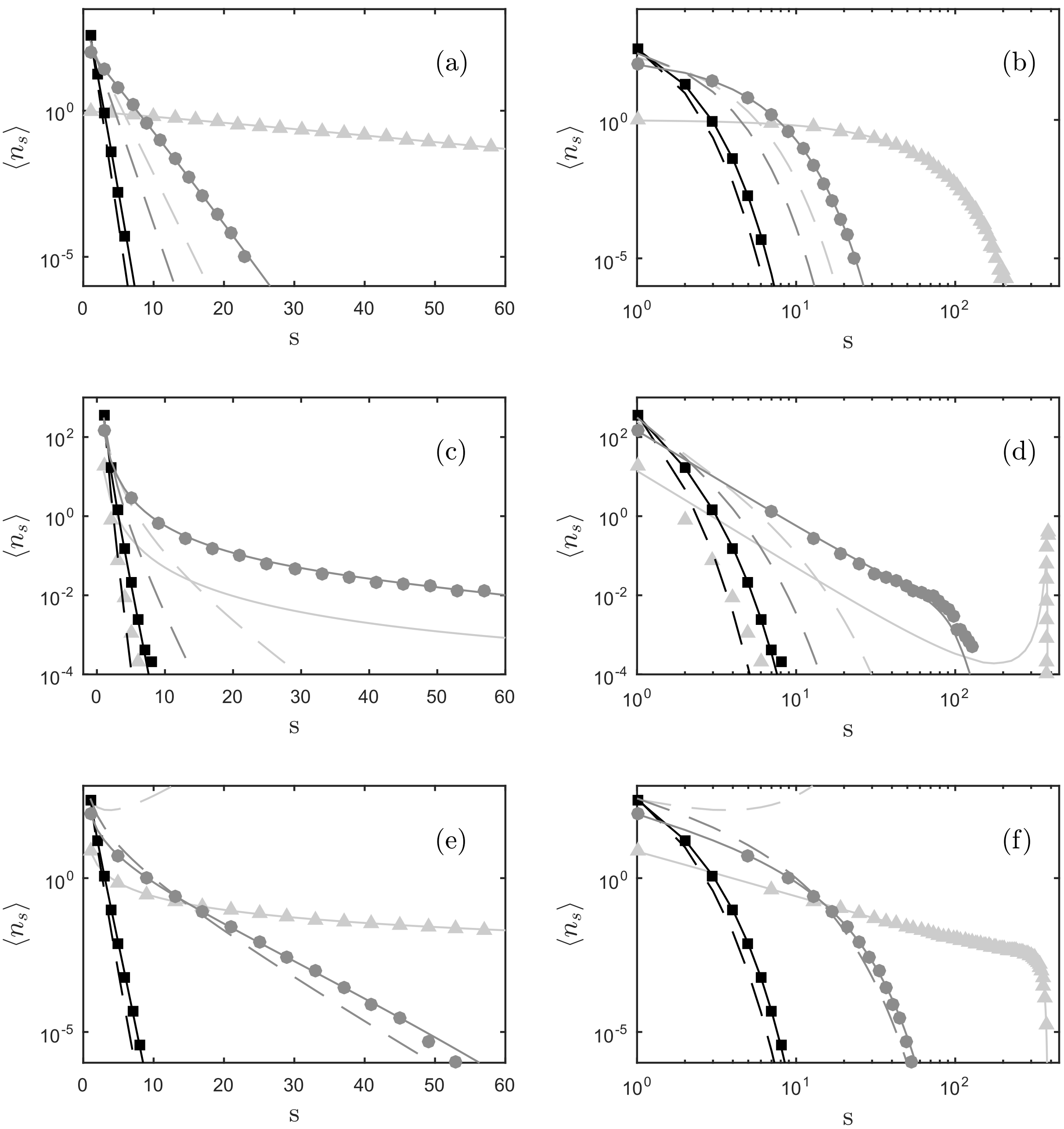

**Fig. 1.** The average number of clusters of size $s$, $\langle n_s \rangle$ for the constant (a, b), multiplicative (c, d) and additive (e, f) kernel. Solid lines represent theoretical predictions based on our combinatorial approach; dashed lines represent the exact solutions of the discrete version of Smoluchowski equations; circles, squares and triangles

correspond to numerical simulations. The coagulation process starts from $N = 400$ monomers. Three cases are presented: $k = 380$ (the beginning of the process, black), $k = 200$ (the half time, dark gray) and $k = 20$ (the very late stage of the process, light gray). For each case $10^6$ independent simulations were performed. **(a, b)** For the constant kernel, as previously shown in [28], the exact theoretical approach perfectly fits numerical simulations while Smoluchowski solutions only approximates the solutions for early stages of the process. **(c, d)** For the multiplicative kernel the compatibility between our combinatorial expressions and simulation are very good for earlier $t$ (higher $k$), however, for later $t$ (lower $k$) predictions do not model the process perfectly and only approximate the exact solution. This deficiency is due to the fact that the multiplicative kernel is one of the gelling kernels. Therefore, at the time of approx. $t \approx N/2$ "percolations occurs" and one giant cluster appears which is not covered by combinatorial expressions. **(e, f)** For the additive kernel, similarly as in the case of constant kernel, our theoretical predictions show excellent compatibility with the results of numerical simulations at all stages of the coagulation process

As can be seen in Fig. 1 the exact approach shows excellent accuracy in case of constant and additive kernels. For the multiplicative kernel, in which the gelation phenomenon occurs, the combinatorial solution is an effective approximation for the time of coagulation process before the system reaches its critical point. The possible explanation of inexactness for the later time of system evolution is that the emergence of one giant cluster during the phase transition in the multiplicative kernel system implies that available ways of creation of a certain cluster are no longer equiprobable which is not covered by the combinatorial expressions.

## VI. Summary

The main achievement of this work is extending the exact combinatorial approach proposed in [28] for additive and multiplicative kernels. The methodology for describing the number of all possible histories of a cluster of given size, $x_g$, has been introduced using recursive equations. This method is general and can be extended for further arbitrary forms of kernels. The correctness of combinatorial solutions has been shown by the comparison with results of numerical simulations for the systems with monodisperse initial conditions. The accuracy of predictions is excellent for the constant and additive kernels for the whole time of the coagulation process. For the multiplicative kernel the combinatorial solutions are not exact due to the gelation, although they give a close approximation of the process for the time before the critical point.

The combinatorial solutions to all three kernels outperforms standard solutions to coagulating systems that arise from the Smoluchowski equations as the latter ones base on the mean–field approximation which is obviously not fulfilled for small systems or for later time of system evolution. The discreteness of combinatorial approach do not diminish the generality of the solutions as it allows to obtain continuous–time results in the way described in [28]. The possible field of use of the results presented here, apart from modeling real systems, are, for instance, percolation phenomena in random networks [30, 31].

Our results can be further developed by analyzing next kernels observed in real systems, e.g. kernels of the form of $\left(i^{\alpha}+j^{\beta}\right)$. Moreover, the robustness to arbitrary initial configuration of clusters other than monodisperse initial conditions applied here needs to be verified.

## VII. Acknowledgments

This work has been supported by the National Science Centre of Poland (Narodowe Centrum Nauki, NCN) under grant no. 2015/18/E/ST2/00560 (A.F. and M. Ł.).